\documentclass{article}
\usepackage{spconf,amsmath,graphicx}
\usepackage{placeins}

\usepackage{enumitem}
\setlist{nosep, leftmargin=14pt}

\usepackage[hyphens]{url}
\usepackage[hidelinks]{hyperref}

\makeatletter
\def\thebibliography#1{%
  \section*{References}
  \addcontentsline{toc}{section}{REFERENCES}
  \list{[\arabic{enumi}]}{%
    \settowidth\labelwidth{[#1]}\leftmargin\labelwidth
    \advance\leftmargin\labelsep
    \usecounter{enumi}%
  }%
  \def\newblock{\hskip .11em plus .33em minus .07em}%
  \sloppy\clubpenalty4000\widowpenalty4000%
  \sfcode`\.=1000\relax%
}

\makeatother

\title{\vspace*{-0.5cm}Large-scale modality-invariant foundation models for brain MRI analysis: Application to lesion segmentation}

\name{%
\shortstack{%
Petros Koutsouvelis$^{1,\dag,*}$,
Matej Gazda$^{2,\dag}$,
Leroy Volmer$^{1}$,
Sina Amirrajab$^{3}$\\
Kamil Barbierik$^{5}$,
Branislav Setlak$^{2}$,
Jakub Gazda$^{4}$,
Peter Drotar$^{2}$%
}%
}

\address{
\textsuperscript{\dag}\small Equal contribution, 
\small \textsuperscript{*} Corresponding author: \protect\texttt{petros.koutsouvelis@maastrichtuniversity.nl} \\
$^{1}$\small Department of Radiation Oncology (Maastro), GROW Research Institute for Oncology and Reproduction,\\
\small Maastricht University Medical Centre+, Maastricht, The Netherlands \\
$^{2}$\small Intelligent Information Systems Lab, Technical University of Kosice, Letna 9, 04201, Kosice, Slovakia \\
$^{3}$ \small The D-Lab, Department of Precision Medicine, GROW - Research Institute for Oncology and Reproduction,\\ 
\small Maastricht University, 6220 MD Maastricht, The Netherlands \\
$^{4}$\small 2nd Department Of Internal Medicine, Pavol Jozef Safarik University and Louis Pasteur University Hospital, Kosice, Slovakia \\
$^{5}$\small SWAI a.s.}
\begin{document}
%
\maketitle
\begin{abstract}
The field of computer vision is undergoing a paradigm shift toward large-scale foundation model pre-training via self-supervised learning (SSL). Leveraging large volumes of unlabeled brain MRI data, such models can learn anatomical priors that improve few-shot performance in diverse neuroimaging tasks. However, most SSL frameworks are tailored to natural images, and their adaptation to capture multi-modal MRI information remains underexplored. This work proposes a modality-invariant representation learning setup and evaluates its effectiveness in stroke and epilepsy lesion segmentation, following large-scale pre-training. Experimental results suggest that despite successful cross-modality alignment, lesion segmentation primarily benefits from preserving fine-grained modality-specific features. Model checkpoints and code are made publicly available.\footnote{\url{https://github.com/BraveDistribution/UMBRA/tree/main}}
\end{abstract}
%



\section{Introduction}
\label{sec:intro}

Fueled by recent advances in self-supervised learning (SSL) and public release of large-scale datasets, the field of medical image analysis is shifting toward pre-training general-purpose foundation models \cite{wald_2025, munk2025}. Compared to training from scratch, this paradigm enables (i) learning representations from unlabeled data that may facilitate the identification of subtle abnormalities, (ii) generalizing to various diagnostic tasks and populations with minimal adaptation. These properties are particularly desirable in brain MRI analyses, that remain constrained by the scarcity of clinical-grade annotated cohorts and heterogeneous acquistion protocols. 

MRI comprises multiple modalities that offer complementary views of the same underlying biological structure. We hypothesize the existence of subject-specific, modality-invariant representations that can be learned during pre-training by mapping input modalities to a shared latent space. Such a model could capture discriminative structural features while mitigating the impact of missing modalities in retrospective cohorts and reducing future acquisition demands. Existing evidence suggests that unifying cross-modality representations can improve brain lesion segmentation, though performed at smaller pre-training scales \cite{chalcroft_2025, tan_2025} or directly during the downstream task \cite{he2024modality}.

This study evaluates the effectiveness of large-scale foundation model pre-training for downstream lesion segmentation, focusing on stroke and epilepsy. A modality-invariant representation learning framework is introduced and integrated with a masked reconstruction objective, given its demonstrated advantages in segmentation tasks \cite{wald_2025}. The contribution of the modality-invariance branch is assessed both independently and in comparison with established SSL approaches. To promote further research, all pretrained foundation models are publicly released.

\begin{figure*}[htbp]
\centering
    \includegraphics[width=0.85\textwidth]{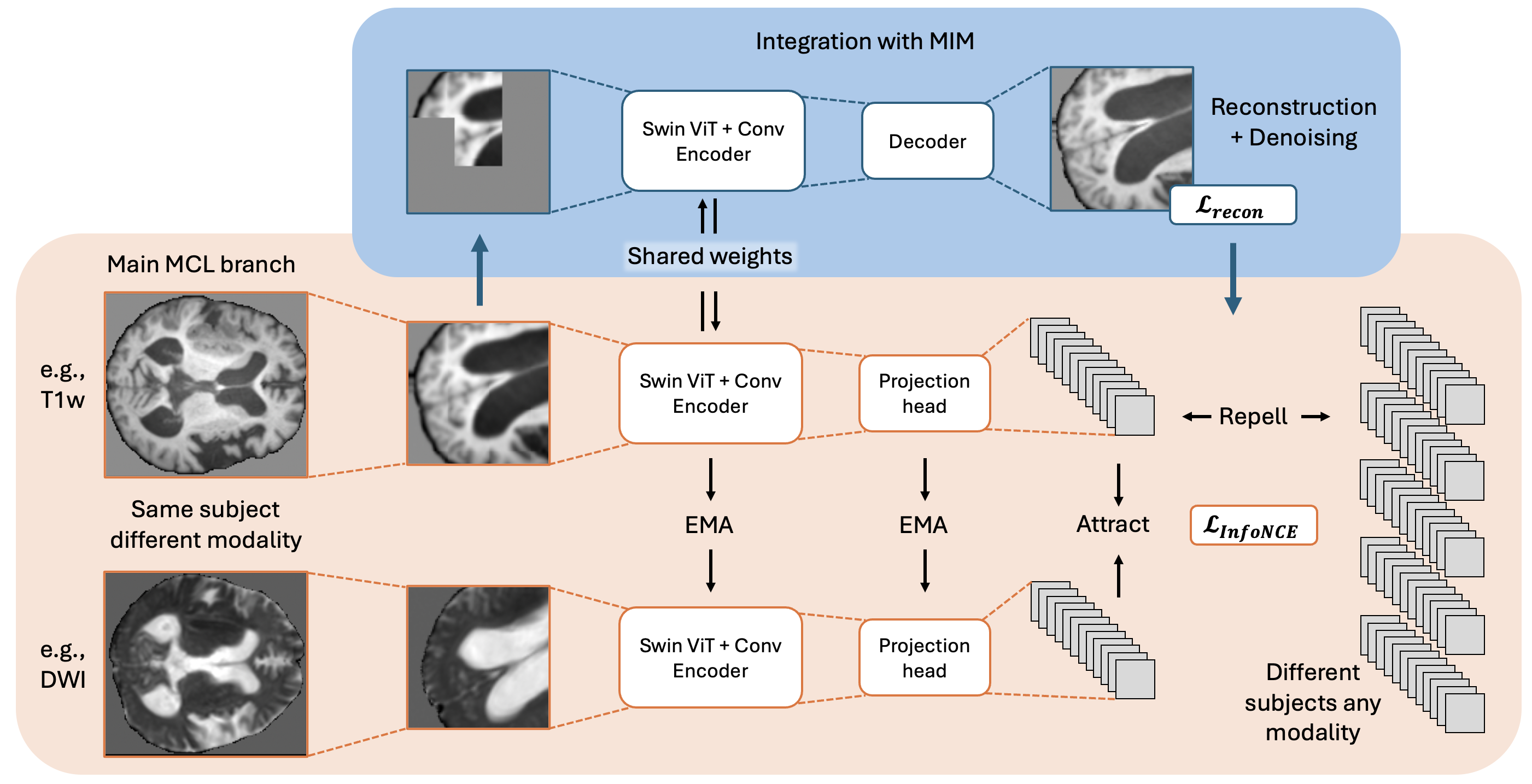}
    \caption{\small High-level schematic of the pre-training setup. \textit{Abbreviations}: EMA = exponential moving average.}
    \label{fig:setup}
\end{figure*}

\section{Methods}

\subsection{Data Curation and Pre-processing}
Pre-training was performed using \textit{FOMO60k} \cite{munk2025}, a large-scale brain MRI dataset containing 60,529 scans and multiple modalities across 11,187 subjects, aggregated from 16 sources. To ensure consistency, all images were skull-stripped with HD-BET \cite{Isensee_2019}, regardless of their original pre-processing pipeline. The scans were then resampled to isotropic 1~mm$^3$ resolution, reoriented to RAS coordinate system, and cropped to the brain bounding box. Intensity values were clipped to the 0.5--99.5\% percentile range and z-score normalized within the brain mask. Images with any spatial dimension smaller than 12~cm were excluded (n = 8,164), as this empirically determined threshold effectively removed scans with missing brain tissue due to registration artifacts.

\subsection{Modality-Invariant Contrastive Learning (MCL)}
We employed a \textit{MoCo v2}-based contrastive learning (CL) setup \cite{chen2020improved}, where co-localized 3D brain MRI patches across all scans in a single acquisition session defined the set of \textit{positive pairs}. Patches from different subjects were considered \textit{negative} irrespective of their location and modality, while same-subject different-session comparisons were excluded due to potential longitudinal shifts. Positive pairs underwent separate affine and intensity augmentations to discourage capturing trival relationships. We compared MCL with the default CL implementation, where positive pairs were defined as augmented views of the same modality.

\subsection{Integration with Masked Image Modeling (MIM)}
Masking was performed at the bottleneck and interpolated back to the patch embedding space to assign masked patches with a learnable token \cite{xie_2022}. This setup ensured no leakage of umasked tokens at any stage of the encoder and compatibility with both convolutional and vision transformer (ViT) layers. The target preceded intensity augmentations to encourage noise-invariant representations. To observe emerging dynamics from the two objectives, the MCL branch was replaced with the default CL as an ablation experiment. A high-level schematic of the setup is depicted in Figure~\ref{fig:setup}.

\subsection{Implementation Details}
A hybrid Swin encoder with residual convolutions \cite{he_2023} was used in all experiments on input volumes of 96 $mm^3$. The queue size in MCL was set to 16,384 and the \textit{InfoNCE} temperature was fixed at $\tau=0.2$ \cite{chen2020improved}. The EMA momentum was linearly increased from 0.996 to 0.999 during the first 20\% of training steps. A lightweight feature pyramid network (FPN) decoder mapped features from the MIM branch to the voxel space. The masking ratio varied uniformly between 0.5 and 0.75. Pre-training was performed for 200k steps using the AdamW optimizer (weight decay 0.01) with a base learning rate of $1\times10^{-4}$ and a batch size of 16. Finetuning continued for 5k steps; the FPN decoder was trained from scratch, while the pre-trained encoder remained frozen for the first 30\% of steps. No hyperparameter tuning was conducted. Segmentation outputs were obtained via sliding-window inference without further post-processing.

\subsection{Evaluation Setup}
Downstream lesion segmentation was evaluated on three cohorts: (i) \textit{ATLAS v2} \cite{liew2022large}, a multi-center dataset of 655 post-stroke cases with $T_1$-weighted images and expert-annotated lesion masks; (ii) \textit{ISLES 2022} \cite{hernandez2022isles}, a two-center dataset of 255 ischemic stroke cases with annotations on DWI and accompanying ADC and FLAIR sequences; and (iii) \textit{FCD BONN} \cite{schuch_2023}, a single-center dataset of 85 focal cortical dysplasia (FCD) patients with manual lesion annotations in FLAIR and additional $T_1$-weighted MRI. All images were rigidly co-registered and pre-processed as \textit{FOMO60k}. 

Unique models were trained per imaging modality and tested on held-out sets using the Dice similarity coefficient (DSC). When feasible, test sets were constructed via center-wise splits; 10 randomly selected centers for \textit{ATLAS v2} (n = 255) and center~\#2 (n = 52) for \textit{ISLES 2022}. \textit{FCD BONN} distinguishes a subset of 28 challenging cases, which served as the held-out set. Given the absence of additional modalities in \textit{ATLAS v2}, the training set (n = 400) was further partitioned into subsets of 100\%, 75\%, 50\%, and 25\% to assess robustness to reduced sample size. Results were compared against training from scratch (i.e., no pre-training) baselines.

\section{Experimental Results}
The proposed modality-invariant pre-training framework was first evaluated for its ability to align cross-modality embeddings, and subsequently for its impact on downstream lesion segmentation performance.

\subsection{Modality Invariance in the Embedding Space}
\begin{figure}[h]
    \includegraphics[width=0.45\textwidth]{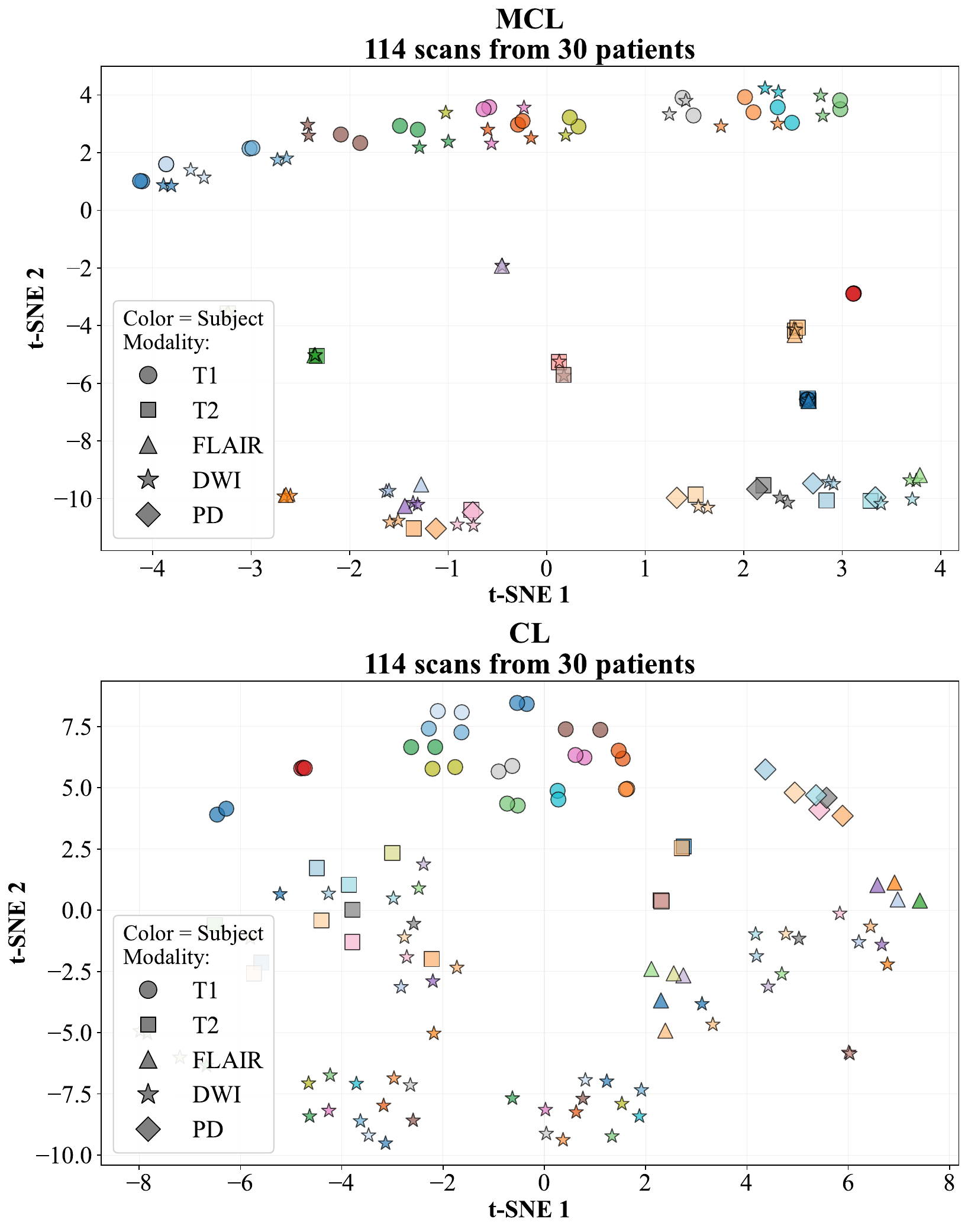}
    \caption{\small tSNE on the embeddings from 30 randomly selected cases of the \textit{FOMO60k} validation set using the MCL and CL models.}
    \label{fig:tsne}
\end{figure}

Cross-modality alignment was measured via the mean cosine distance between encoder embeddings from different acquisitions of the same anatomical region. Given the centrally-cropped patch of each subject/session combination in a 2\% held-out validation set from \textit{FOMO-60k}, modality-invariant pre-training demonstrated effective cross-modality alignment (MCL: $0.0066 \pm 0.0104$, MCL + MIM: $0.0068 \pm 0.0175$) compared to the default counterparts (CL: $0.5995 \pm 0.1576$, CL + MIM: $0.5028 \pm 0.1592$). This is further illustrated by the t-SNE plots in Figure~\ref{fig:tsne}, where cross-modality embeddings appear considerably more clustered. Cross-subject separability in $T_1$-weighted inputs remains relatively consistent across frameworks (MCL: $0.6094 \pm 0.2500$, MCL + MIM: $0.6232 \pm 0.2732$, CL: $0.6540 \pm 0.2023$, CL + MIM: $0.6146 \pm 0.2198$), suggesting no collapse of discriminative features.

\subsection{Lesion Segmentation Performance}
\begin{figure}[h!]
    \includegraphics[width=0.47\textwidth]{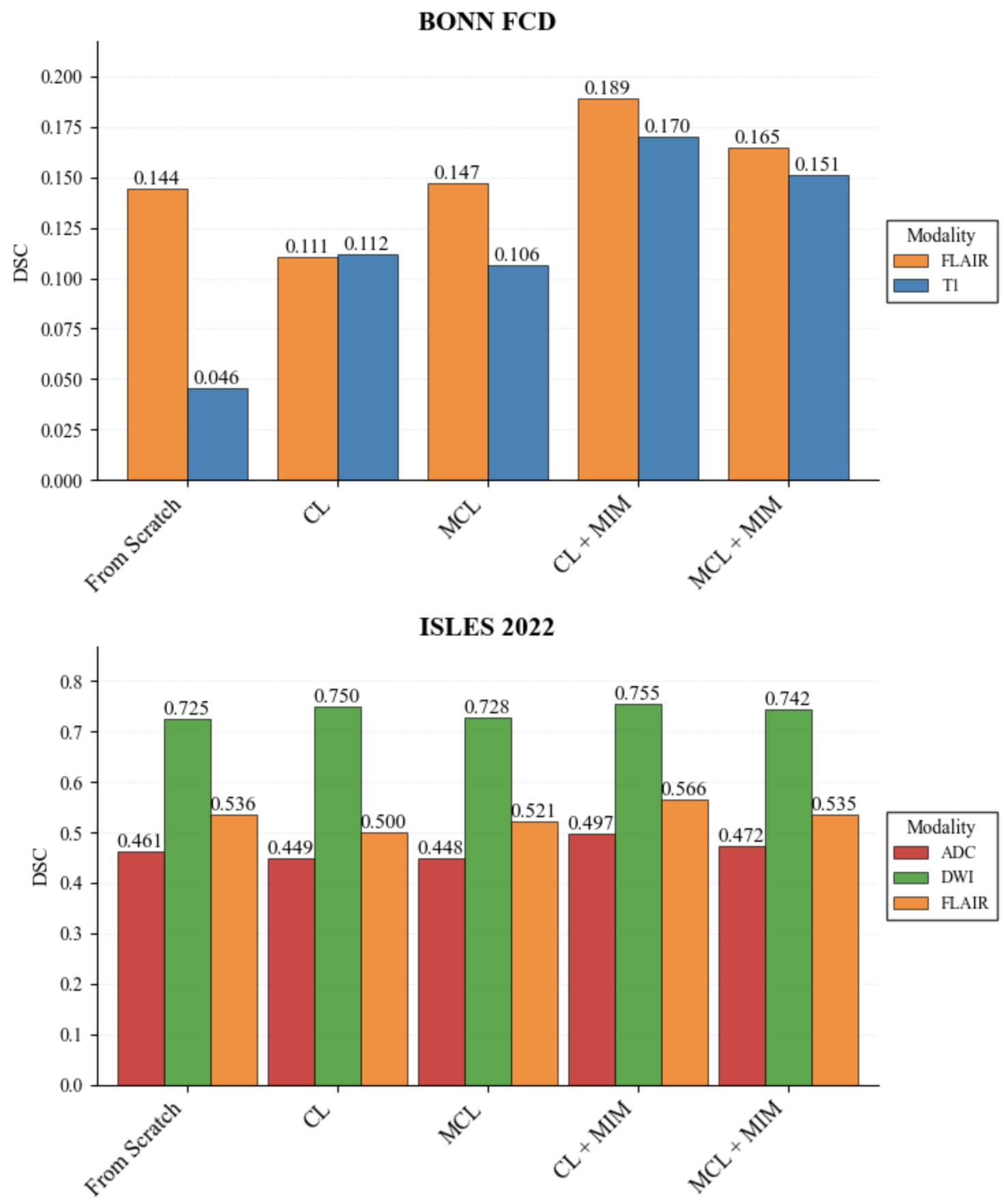}
    \caption{\small Segmentation DSC in the \textit{BONN FCD} and \textit{ISLES 2022} datasets for each modality and pre-training method.}
    \label{fig:fcd_and_isles}
\end{figure}

Segmentation results across all datasets, modalities, and pre-training  methods are summarized in Table~\ref{tab:atlas} and Figure~\ref{fig:fcd_and_isles}. Overall, aligning cross-modality representations during pre-training did not steer the models into capturing richer spatial information, highlighted by the comparable performance between MCL and default CL. Instead, performance was mainly driven by the presence of a MIM branch, due to the property of the reconstruction objective to encourage highly-localized embeddings. 

Integrating MCL with MIM appeared to deteriorate per-modality performance compared to the default CL with MIM, except in few-shot ablations. This could be attributed to the MCL objective limiting the effectiveness of MIM by (i) yielding higher gradient norms and consequently dominating the optimization process, and (ii) eliminating modality-specific texture features that are necessary to reconstruction. The use of a higher $\tau$ value in \textit{InfoNCE}, combined with a token-level modality-agnostic reconstruction objective, such as \textit{iBOT} \cite{zhou_2022}, could circumvent these limitations and surpass the default baseline, while preserving modality invariance. 

\begin{table}[ht]
\centering
\caption{\small Segmentation DSC with respect to \% of training data used in the \textit{ATLAS v2} dataset.}
\vspace{5pt}
\label{tab:atlas}
\begin{tabular}{lrrrr}
\hline
\textbf{Method} & \textbf{100\%} & \textbf{75\%} & \textbf{50\%} & \textbf{25\%} \\
\hline
From Scratch & 0.4188 & 0.4339 & 0.3991 & 0.3918 \\
CL & 0.4678 & 0.4648 & 0.4552 & 0.4377 \\
MCL & 0.4693 & 0.4572 & \textbf{0.4892} & 0.4420 \\
CL + MIM & \textbf{0.4936} & 0.4872 & 0.4643 & 0.4530 \\
MCL + MIM & 0.4782 & \textbf{0.4893}& 0.4576 & \textbf{0.4564}\\
\hline
\end{tabular}
\end{table}

\subsection{Modality Invariance in Downstream Performance}

Results in Figure~\ref{fig:fcd_and_isles} indicate a strong coupling between imaging modality and downstream performance, independent of the pre-training strategy. The highest-scoring modalities were FLAIR for FCD and DWI for stroke lesions, which is concordant with their diagnostic value in clinical practice. This pattern appears unconfounded by the representativeness of the pre-training cohort, as it persists even in models trained from scratch. However, it could be partly attributable to the ground truth annotations being created on the same modalities. 

Interestingly, pre-trained MCL and MCL + MIM encoders yielded more aligned cross-modality embeddings in the downstream test sets (e.g., in \textit{ISLES 2022} - MCL: $0.1231 \pm 0.1126$, MCL + MIM: $0.1312 \pm 0.1503$, CL: $0.6291 \pm 0.0816$, CL + MIM: $0.6296 \pm 0.0804$), yet segmentation performance showed similar variation across modalities. This suggests that (i) modality-invariant representations may emphasize global, patient-level anatomical features rather than localized pathological abnormalities, and (ii) each modality captures distinct tissue characteristics that cannot be inferred from other modalities and are hence essential to segmentation. In this regard, the proposed pre-training method could prove more effective for global classification or regression tasks, such as brain age prediction, whereas segmentation studies would benefit from preserving discriminative, modality-specific features and, when feasible, unifying representations through multi-modal fusion.

\section{Conclusion}
This study proposed a modality-invariant self-supervised learning framework and evaluated its effectiveness for lesion segmentation following large-scale pre-training. Experimental results showed that while aligning intra-patient cross-modality representations was feasible, it neither improved per-modality lesion segmentation performance nor reduced dependency on specific modalities. While performance could be improved by optimizing the pre-training setup, the latter may not be feasible, as accurate lesion segmentation relies on fine-grained spatial detail that remains closely tied to modality-specific contrasts and pathology. Future work should investigate the role of modality invariance in global classification or regression tasks, where such representations may prove more beneficial.

\section{Acknowledgments}
\label{sec:acknowledgments}
This work is funded by the EU NextGenerationEU through the Recovery and Resilience Plan for Slovakia under the project No. 09I03-03-V04-00394 and used the Dutch national e-infrastructure with the support of the SURF Cooperative using grant no. EINF-13271.

\small
\bibliographystyle{IEEEtran}
\bibliography{strings}

@article{chen2020improved,
  title={Improved baselines with momentum contrastive learning},
  author={Xinlei Chen and others},
  journal={arXiv preprint arXiv:2003.04297},
  year={2020}
}

@ARTICLE{schuch_2023,
  title    = "An open presurgery {MRI} dataset of people with epilepsy and
              focal cortical dysplasia type {II}",
  author   = "Schuch, Fabiane and others",
  journal  = "Scientific Data",
  volume   =  10,
  number   =  1,
  pages    = "475",
  month    =  jul,
  year     =  2023,
}

@article{Isensee_2019,
    title={Automated brain extraction of multisequence MRI using artificial neural networks},
    volume={40},
    DOI={10.1002/hbm.24750},
    number={17},
    journal={Human Brain Mapping},
    publisher={Wiley},
    author={Isensee, Fabian and others},
    year={2019},
    month=aug, pages={4952–4964}
}

@misc{wald_2025,
      title={An OpenMind for 3D medical vision self-supervised learning}, 
      author={Tassilo Wald and others},
      year={2025},
      eprint={2412.17041},
      archivePrefix={arXiv},
      primaryClass={cs.CV},
}

@misc{xie_2022,
      title={SimMIM: A Simple Framework for Masked Image Modeling}, 
      author={Zhenda Xie and others},
      year={2022},
      eprint={2111.09886},
      archivePrefix={arXiv},
      primaryClass={cs.CV},
}

@InProceedings{he_2023,
    author="He, Yufan
    and others",
    title="SwinUNETR-V2: Stronger Swin Transformers with Stagewise Convolutions for 3D Medical Image Segmentation",
    booktitle="Proc. MICCAI 2023",
    year="2023",
    publisher="Springer Nature Switzerland",
    address="Cham",
    pages="416--426",
}

@misc{zhou_2022,
      title={{iBOT}: Image BERT Pre-Training with Online Tokenizer}, 
      author={Jinghao Zhou and others},
      year={2022},
      eprint={2111.07832},
      archivePrefix={arXiv},
      primaryClass={cs.CV},
}

@misc{chalcroft_2025,
      title={Unified 3D MRI Representations via Sequence-Invariant Contrastive Learning}, 
      author={Liam Chalcroft and others},
      year={2025},
      eprint={2501.12057},
      archivePrefix={arXiv},
      primaryClass={cs.CV},
}

@misc{tan_2025,
      title={Towards a Universal 3D Medical Multi-modality Generalization via Learning Personalized Invariant Representation}, 
      author={Zhaorui Tan and others},
      year={2025},
      eprint={2411.06106},
      archivePrefix={arXiv},
      primaryClass={cs.CV},
}

@misc{munk2025,
    title={A large-scale heterogeneous 3D magnetic resonance brain imaging dataset for self-supervised learning},
    author={Asbjørn Munk and others},
    year={2025},
    eprint={2506.14432},
}

@article{liew2022large,
  title={A large, curated, open-source stroke neuroimaging dataset to improve lesion segmentation algorithms},
  author={Liew, Sook-Lei and others},
  journal={Scientific data},
  volume={9},
  number={1},
  pages={320},
  year={2022},
  publisher={Nature Publishing Group UK London}
}

@article{hernandez2022isles,
  title={ISLES 2022: A multi-center magnetic resonance imaging stroke lesion segmentation dataset},
  author={Hernandez Petzsche, Moritz R and others},
  journal={Scientific data},
  volume={9},
  number={1},
  pages={762},
  year={2022},
  publisher={Nature Publishing Group UK London}
}

@inproceedings{he2024modality,
  title={Modality-agnostic learning for medical image segmentation using multi-modality self-distillation},
  author={He, Qisheng and others},
  booktitle={Proc. ISBI 2024},
  pages={1--5},
  year={2024},
  organization={IEEE}
}

\end{document}